# IoT2Vec: Identification of Similar IoT Devices via Activity Footprints

Kushal Singla, Joy Bose
Samsung R&D Institute
Bangalore, India
{kushal.s, joy.bose}@samsung.com

*Abstract*— We consider a smart home or smart office environment with a number of IoT devices connected and passing data between one another. The footprints of the data transferred can provide valuable information about the devices, which can be used to (a) identify the IoT devices and (b) in case of failure, to identify the correct replacements for these devices. In this paper, we generate the embeddings for IoT devices in a smart home using Word2Vec, and explore the possibility of having a similar concept for IoT devices, aka IoT2Vec. These embeddings can be used in a number of ways, such as to find similar devices in an IoT device store, or as a signature of each type of IoT device. We show results of a feasibility study on the CASAS dataset of IoT device activity logs, using our method to identify the patterns in embeddings of various types of IoT devices in a household.

## I.  INTRODUCTION

The usage of IoT devices is increasing exponentially, with billions of such devices being used on a daily basis. It would be useful to have a system to identify these IoT devices based on their patterns of usage in one or more given households. Such identification of IoT devices on the basis of usage might be useful in, for example, identifying the correct replacements for such devices in case of damage or in knowing which IoT device to buy based on usage requirements. The popular Word2Vec model [1] provides ways to generate word embeddings on the basis of usage of the words in a given dataset of documents. Using a similar concept, we attempt to create embeddings for IoT devices on the basis of their usage, using data obtained from the usage logs. Such a model we call IoT2Vec.

In this paper, we describe the generation of embeddings using some publicly available IoT datasets and describe some principles how this can be adapted for different IoT device usage data. We create a model to take the devices usage data as input, create embeddings for the devices and identify a new device of the same type based on similar usage data.

The rest of this paper is organized as follows: in the following section we survey approaches to current approaches related to identifying similar IoT devices. In Section 3, we discuss several use cases of the concept, which serve as a motivation for having a separate embedding mechanism for IoT devices. Section 4 describes the theory and method which we use to generate the embeddings. Section 5 gives the results of our method applied on a public IoT dataset. Section 6 suggests some further questions to analyze in order to learn from activity patterns of various types of IoT sensors and devices. Section 7 concludes the paper.

## II.  RELATED WORK

There are a few instances of related work in the area of applying machine learning to find similar IoT devices.

Xu [2], in his PhD thesis, proposes a system for searching for and finding similar IoT devices as a result of user queries, based on a similarity measures based on the semantic and other properties of the IoT objects such as the object location.

Kang [3] suggested various methods to identify correlations between IoT devices, including attributes such as location, usage count, sensor list, service name etc. They also suggested using Word2Vec model to calculate the adjacency between IoT devices. However, they did not provide concrete details on exactly how the vectors would be calculated and the issues involved when working with real datasets.

Tian [4] mentioned a mechanism to automatically collect security related information from an IoT app. It included a feature to analyze app descriptions using NLP techniques and extract capability information for various IoT sensors.

Palit [5] mentioned a system to identify IoT resource requirements such as sensor accesses from service descriptions. For example, their method included determining which IoT devices would be required to perform a specific task such as tracking a device. They used NLP techniques to parse the Android app descriptions to determine which sensors were required to accomplish the tasks mentioned in the descriptions.

Hong [6] used similarity measures between IoT devices to provide context aware services to users. Their main aim was to recognize a place based on the IoT devices kept in that location, using the hypothesis that similar devices are kept in similar locations. For example, a living room is expected to contain a TV and so on.

Truong [7] propose a method for searching similar IoT sensors, computing a similarity score based on fuzzy sets. In their approach, the fuzzy sets are constructed based on a range of IoT sensor attributes such as location or temperature range. They determined that sensors in similar locations tended to be similar.

In this paper, we take a different approach, focusing on IoT sensor or device usage patterns create the word embeddings and identify the type of IoT device. Our approach has the advantage of not needing any prior information or assumptions about the IoT devices. So it can be used in a wider variety of potential use cases. On the other hand, we are solving a harder problem as patterns may or may not be easy to be discerned.

## III. MOTIVATION

As we can see from a survey of the above related work, there can be a number of applications to finding similar IoT devices. One obvious application can be to make a search function to search for similar devices in the vicinity. This was one of the applications studied by Xu [2]. Another application was that given a requirement, one could determine the combination of IoT devices to fulfill that requirement. This was done using the rule based device similarity calculations. Given a certain workflow of functionality, we can find the devices to fulfil the functionality. Our approach of determining word embeddings for IoT devices can also fulfill a similar use case.

Another motivation can be: defective devices can be replaced based on their function. If we know the footprint of the IoT device, we can identify which other device is best suited to replace it, such as when we are buying from an IoT store.

Another common application can be to build a location classifier based on IoT devices in that location. For example, given that a bar can use dim lights, it is likely that another bar will have similar lights. So knowing the IoT devices and their footprint in a given location, we can identify the type of location. Hong [6] concentrated on this requirement. Another use of finding similar IoT devices is to search for sensors which are similar. This is the idea explored by Truong [7].

Hence, any of the above requirements can serve as a motivation for our approach.

In the following section, we describe the basics of our approach and our method for generating word embeddings for IoT devices.

## IV. IOT2VEC THEORY

To fulfill our aim of discovering word embeddings corresponding to IoT devices from the device activity, we postulate the following:

### A. Lemma 1

Most IoT devices are used in certain patterns that repeat over time. Similar kinds of IoT devices will have similar activity footprints.

For example: in multiple homes if a person is performing a similar activity, similar kinds of IoT devices will be active around the same time. For example, if a person goes to a kitchen to cook some food, sensors in the kitchen such as microwave, fridge etc. will get activated. Since kitchens in any two apartments are expected to have similar devices, their patterns of usage will be similar. For example:

Home 1: DF1 (Fridge Door) OPEN in Kitchen → LB1 (Lightbulb) ON Kitchen → DM1 (Microwave Door) ON Kitchen

Home 2: LB2 (Lightbulb) ON Living Room → DF2 (Fridge Door) OPEN Kitchen → DM1 (Microwave Door) ON Kitchen → AC2 ON Living Room

### B. Lemma 2

The IoT device type can be identified by the pattern of usage.

For example, the AC would have a different pattern of usage than, say, a TV. The AC might be turned ON during the nighttime when a person is planning to go to sleep. The TV might be ON during the daytime or when the person is awake. A burglar alarm would be activated mainly when the person is preparing to leave the house, and so on. Hence, generalizing from the activity of similar IoT devices, we might be able to identify a device or sensor from its activity pattern.

### C. Lemma 3

The time of usage and location of devices also carries useful information.

For example, if a person is preparing food to eat, they might open both the fridge door and the microwave door around the same times. This kind of pattern (door opening of fridge and microwave) might indicate the presence of these devices.

### D. Lemma 4

A model can be trained to encode this pattern as word embeddings, which will help to identify the IoT device that has similar patterns.

For example, two houses might have similar TV, Fridge and AC usage patterns. Recognizing these similar patterns can help to create embeddings for these devices using tools such as Word2Vec or Glove.

### E. Method to create word embeddings for IoT devices

Using the previously mentioned postulates, we define some steps to generate and analyze the word embeddings from IoT device sensor logs.

Our method includes the following steps:

1. Filter out the IoT sensors whose data is not meaningful or we cannot make sense of the data.

2. Examine the activity data of the selected sensors to see whether it shows meaningful activity or actions.

3. Extract only the values where the sensor state is in transition (e.g. ON to OFF or OFF to ON).

4. Build a session of the sensor values (similar to sentence in NLP domain) by choosing a session gap. Session gap is the gap of time where we construct the boundaries of each session. Within a session, we only consider the order of firings of different devices or sen-sors. The exact time gap between firings within a session is ignored.

5. Once the sessions are defined, we treat each session as a sentence and the device Id as a word. Each sentence will contain a sequence of IoT device Ids such as (M008, M009, D010) which is the order of firings of the devices within the session.

6. Train the Iot2Vec model using the session data extracted from the dataset. The input to the training model is the document comprising of the created sentences in the previous step. The output of the model is the embedding vector for each type of sensor or device. We can select a certain dimension, such as 100, for the size of the vector embeddings.

7. Compute the similarity between the vector embeddings of each sensor/device with the other sensors. Furthermore, we perform dimensionality reduction and construct a t-SNE plot for easier visualization of the sensor activities in terms of contextual similarity, i.e. which IoT devices or sensors are being activated together.

8. Visually examine the t-SNE plots to detect patterns of similarity in the activity data for each type of IoT device or sensor with other sensors.

Following the above steps, the embeddings vector of a given device or sensor type can be generated from its activity logs.

TABLE I.  ALGORITHM TO IDENTIFY DEVICE TYPE FROM ACTIVITY LOGS

ALGORITHM: Algorithm to identify device type of unknown IoT device

***Input***: Stored device embeddings for different device or sensor types $D_1 (E_1), D_2 (E_2)$ etc.
***Output***: Device type of a new device $D_i$ given its usage data
1. Generate embeddings vector $E_j$ from the usage data of the new device $D_j$
2. Compute the similarity of the embedding vector $E_j$ with each of the stored embedding vectors $E_1, E_2 \ldots$
3. Find the device $D_i$ whose similarity value of the embedding vector $E_i$ with $E_j$ is highest and above a threshold
4. Define the device type of $D_j$ as equal to the device type of $D_i$
5. *exit*: end procedure

### F. Method to identify the device type of a new or unknown IoT devices from its usage logs

The table 1 shows the algorithm for identifying the device type of a new or unknown IoT sensor or device from its activity logs, once we have stored device embeddings of a set of devices. The principle is to generate the embeddings vector for the new device, and determine which of the stored embeddings is closest to the generated embeddings vector.

## V. EXPERIMENTAL DETAILS AND RESULTS

For our experiments to validate our method and to explore the possibility of generating embeddings for IoT devices, we needed a dataset that would provide us with data from multiple IoT devices in the same locations over a period of time.

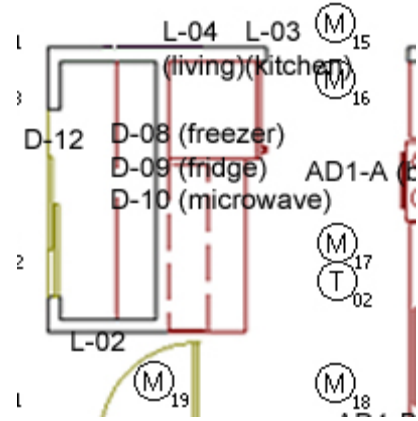

Fig. 1. Extract from the layout of a room in the CASAS Kyoto dataset [9], showing how some motion sensors (beginning with M) and doors (beginning with D) are located close together in the kitchen.

For our purpose, we tried a number of candidate datasets [8-10] and finally chose a dataset 20 from the Kyoto dataset list of CASAS [10, 11]. This dataset has 2 years' worth of data from a household consisting of two residents, with various IoT devices including motion sensors, doors (fridge, freezer, and microwave), shelves etc. Each data item consists of the following fields: time, sensor name, sensor state.

Fig. 1 shows an extract from a layout of a room in a house in the CASAS Kyoto dataset [10, 11].

We then created word embeddings for various devices in the dataset for which we have data.

We then used this embeddings to identify the devices.

We analyzed the dataset in Spark and used Word2Vec to find patterns in the data.

During the preprocessing step, we ignored the light sensor, gyro sensor and a few others, since they were firing without any discernible patterns.

We selected the following sensors for analysis: Motion sensor, door sensor, item sensor, shake sensor, fan sensor, experimental switch.

We then obtained a sequence of sensor states, belonging to multiple sensors, ordered by time. We ignored the actual time of sensor state change and only noted the sequence. Our objective, as mentioned earlier, was to determine the similarity between different sensors on the basis of their activity.

Fig. 2 shows the plot of the sensor states, using the t-SNE model [12] with PCA for dimensionality reduction. We made the t-SNE plots for the devices for three different choices of session gaps (the gap of time to identify a session for a given device): 10 seconds, 60 seconds and 600 seconds. As we can see, a few of the devices show clusters of activity. On examina-tion, such devices may be located in the same regions in the home, although that is not always the case. Moreover, we see that as we vary the session gap, different kinds of devices seem to cluster together.

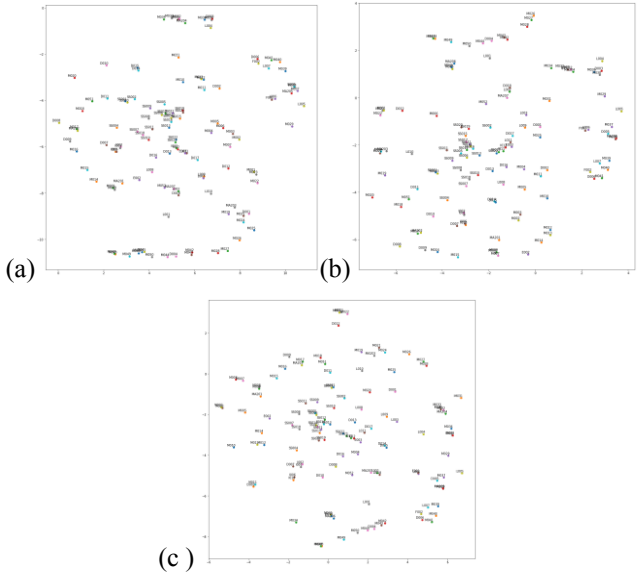

Fig. 2. t-SNE plots of the word embeddings on the Kyoto dataset of CASAS [10], using (a) 10 seconds, (b) 60 seconds and (c) 600 seconds as gap to identify a session. We can see that some kinds of clusters can be discerned.

In the following subsections, we analyze a few trends of device activity based on various parameters.

### A. Trends in IoT device activations for a given value of session gap

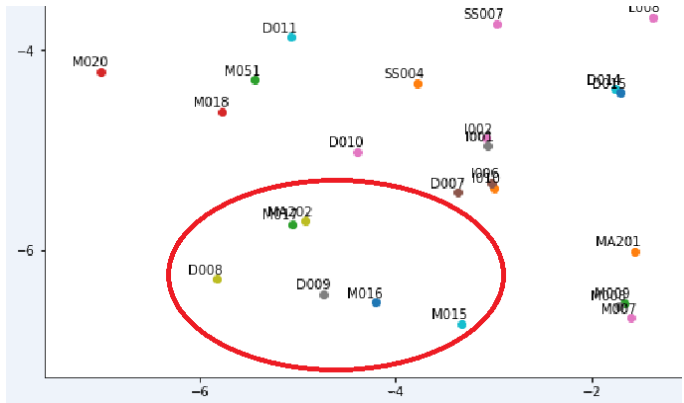

Fig. 3. Extract from the t-SNE plot of the activity of sensors from the CASAS Kyoto dataset [10], using 60 seconds as the interval, showing the contextual proximity of the motion sensors close to the kitchen (M015, M016) and door sensors of fridge (D009) and freezer (D008).

In this subsection, we attempt to find some trends in device activity keeping the session time gap constant. We choose a session time gap of 60 seconds, meaning that we define device activity within 60 seconds as belonging to the same session.

Fig. 3 shows the t-SNE plot of the activity of various sensors in the CASAS Kyoto dataset using 60 seconds as the interval to identify a session.

In our chosen dataset [9, 10], the device D008 is a door sensor corresponding to a freezer door, where the freezer is located in the kitchen. The similarity between vector embeddings that we obtained for this D008 sensor for the 60 second session gap is as below:

```
D008 [('M017', 0.49945521354675293), ('M016',
0.48164984583854675), ('MA202', 0.4487079977989197),
('M018', 0.4332207143306732), ('D009',
0.416538894176483150), ('D015', 0.3721662163734436),
('M015', 0.3238069415092468), ('M051',
0.2985246777534485), ('D010', 0.2684941589832306),
('D014', 0.24952027201652527)]
```

From the above similarity between vector embeddings, we can derive the same conclusion, that door sensor D008 is close to motion sensors M017 and M016 which are in close proximity.

Examining the figure 3, we see that door sensor D008 (freezer, located in the kitchen) comes contextually close to motion sensors M015 and M016, located close to the kitchen, and the door sensor D009 of the fridge, also located in the kitchen. Looking at the layout of the house in fig. 2, we see that the sensors D008 and D009 are located in the kitchen and motion sensors M015, M016, M017 are located close to the kitchen. We can explain their contextual similarity as follows: when a person comes into the kitchen, they would activate the motion sensors close to the kitchen, and would open the fridge and freezer to get or make some food.

Based on this, we conclude that it is feasible to identify IoT devices based on their contextual similarity.

### B. Trends in IoT device activations for varying values of the session gap

In the second analysis, we attempt to find whether the IoT device activity across different session gaps shows any significance. For example, it is possible that when we choose a small gap such as 10 seconds, device A and B are close, however when we increase the time gap to say 600 seconds, device A and C are close. So we visually inspect the T-SNE plots [12] of the activity data to see if that can indeed be the case, or whether similar devices always cluster together.

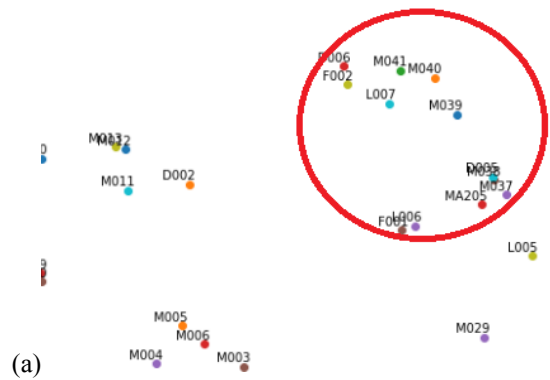

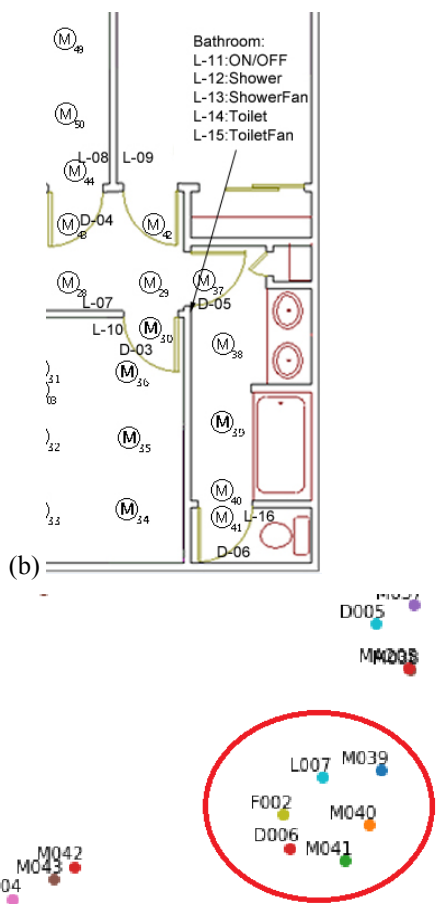

Fig. 4. (a) Extract from the t-SNE plot with 10 seconds as the interval, and (b) location of sensors near the toilet. (c) Sensor activity for 600 seconds gap. Here, sensors located near the toilet M038, M039, M040, M041, D006, D005 show similar patterns of activity across session gaps.

Fig. 4 shows the device activity and sensor locations near the toilet area for 10 seconds session gap. On visual inspection, we can see that motion sensors M039, M040, M041, D006, D005 etc. show correlated activity patterns. This is also confirmed from a look at the similarity measure of distance, where the vector embeddings for these sensors show closest Euclidean similarity between each other. Fig. 5 shows the same sensors activity for a 600 seconds gap for a session. We see that the same sensors that were active together for a 10 second session gap are also active for a 600 seconds session gap.

Hence, we conclude that for this choice of sensors, the proximity of location (all these sensors are in the toilet area) translates into contextual proximity as well, regardless of session intervals. This could be because whenever someone uses the toilet, the motion sensors and door would always be triggered together. However, for a different choice of sensors, this might not be the case and the timer could be a factor in deciding which sensors trigger together.

## VI. Further Questions to Analyze

In our previous section, we found that certain patterns of IoT device activity can be discerned, on the basis of which it is feasible to generate word embeddings for IoT devices based on their usage. A few further questions for analysis can be as below:

- What the differences are of trends in activity for IoT sensors kept in single person households Vs multiple person households?
- Which IoT sensors or devices generate useful data and which ones give noise? For example, how useful is the data generated by the shake sensor vs motion sensors vs door sensors.
- How much session gap is optimal for each type of sensor?
- What additional information about contextual similarities of the device activations do we obtain by varying the dimensionality of the vector embeddings and other input parameters?

## VII. Conclusion

In this paper, we have proposed a method to generate word embeddings for IoT devices, based on their usage patterns. We showed that IoT devices in similar areas in a given household can be found to have similar usage patterns. Thus, it is feasible to recognize IoT devices on the basis of the embeddings. In future, we plan to investigate multiple datasets, generalize our approach and build a number of use cases based on this. We also plan to focus on activity generated by smart IoT devices such as fridge and TV, to get higher level understandings of the patterns of user's tasks.